\documentstyle[epsfig]{l-aa}

\begin{document}

\thesaurus{09(02.03.3; 04.01.2; 06.16.2; 08.01.3; 08.12.1)}  

\title{Fe\,{\sc i} line shifts in the optical spectrum of the Sun}

\author{C.~Allende Prieto \and R. J.~Garc\'{\i}a L\'opez}

\offprints{C.~Allende Prieto (callende@iac.es)}

\institute{Instituto de Astrof\'\i sica de Canarias, E-38200 La Laguna, 
           Tenerife, Spain}                     
  
\date{Received date, ; accepted date, }

\maketitle

\markboth{Allende Prieto \& Garc\'{\i}a L\'opez: Fe\,{\sc i} 
line shifts in the optical spectrum of the Sun}{}

\begin{abstract} 

New improvements in the measurement of both the optical solar spectrum and
laboratory wavelengths for lines of neutral iron are combined to extract
central wavelength shifts for 1446 lines observed in the Sun. This provides
the largest available database of accurate solar wavelengths useful as a
reference for comparison with other solar-type stars. It is shown how the velocity shifts correlate with line strength, approaching a constant value, close to zero, for lines with equivalent widths larger than 200 m\AA.


\end{abstract}

\section{Introduction}
\label{sec1}

Convective motions in the photosphere are nowadays believed to be the main
contributors to observed line asymmetries and  shifts in the optical 
spectra of the Sun and solar-type stars. Several studies have been devoted to
the analysis  and classification of these observational features in the Sun (e.g. Dravins, Lindegren, \& Nordlund 1981, hereafter DLN; Balthasar
1984). Some researchers, extending this work to other stars, have seen how the
difficulties grow, not only due to the lack of photons but also to the
impossibility of accurately removing two major effects: gravitational  shifts
and the radial velocity of the star,  which in turn do not allow the setting up of an
absolute  velocity scale (e.g. Gray 1982; Dravins 1987a,b). However, 
absolute line shifts should be included in spectral syntheses and
inversion codes that take velocity patterns into account.  Line asymmetries  and shifts establish a footprint of the dynamics of convection,
thereby imposing a major constraint to the theoretical modelling.

The increasing resolving power of the spectrographs and the
improvements of detectors have made systematic high resolution
spectroscopic observations possible over wide spectral ranges.  Recent
studies (e.g. Allende Prieto et al. 1995)  have extended very
high-quality  spectroscopic observations beyond the domain of the
brightest stars. In the solar case,  optical, IR and UV atlases are
available from modern observations with very high signal-to-noise ratio
and spectral resolution.  For the optical spectrum of the Sun seen as a
star, the {\it Solar Flux Atlas from 296 to 1300 nm}  (Kurucz et al. 1984,  referred to here as the FTS flux spectrum)  has been the most
extensively used. Partially  from the same data set (flux), partially from a second set of Kitt Peak FTS spectra (disc-centre intensity), H. Neckel prepared
the {\it Spectral Atlas of Solar Absolute Disk-Averaged and Disk-Center
Intensity from 3290 to 12510 \AA} (Brault \& Neckel 1987; for details see Neckel 1994).  We shall refer to the included
disc-centre spectrum as the FTS disc-centre spectrum.  These atlases
achieve signal-to-noise ratios of about 2500 and a resolving power
$\lambda /\Delta\lambda \sim 400000$.

The set of lines with wavelengths accurate enough
to be useful in this context has  been significantly enlarged  by the
work of Nave et al. (1994), who measured and identified 9501 Fe\,{\sc
i} lines using Fourier transform spectrometers at NSO\footnote{National Solar
Observatory, Tucson, Arizona, USA},  Imperial
College (London, UK), and NIST\footnote{National Institute of Standards
and Technology, Gaithersburg, Maryland, USA.}, from 0.17 to 5
$\mu$m.  In the optical range, they claim an uncertainty below 1 m\AA\ for many lines, and two orders of magnitude larger for the worst cases.

We have used these relatively new atlases, as well as the more traditional
Li\`ege Atlas (Delbouille, Neven \& Roland 1973), to measure central wavelengths
for a large subsample of the neutral iron lines included in the line list by
Th\`evenin (1989, 1990). His compilation joins those lines classified by
Moore, Minnaert \& Houtgast (1966) as one-blended or unblended, and it was taken as a starting
point since the determination of accurate line centres does not require such
a clean profile as is needed to measure line asymmetries. This information
is combined with the rest air wavelengths from Nave et al. (1994) to deduce
the displacements of the solar lines.

\begin{figure*}
\begin{center}
\mbox{\epsfig{file=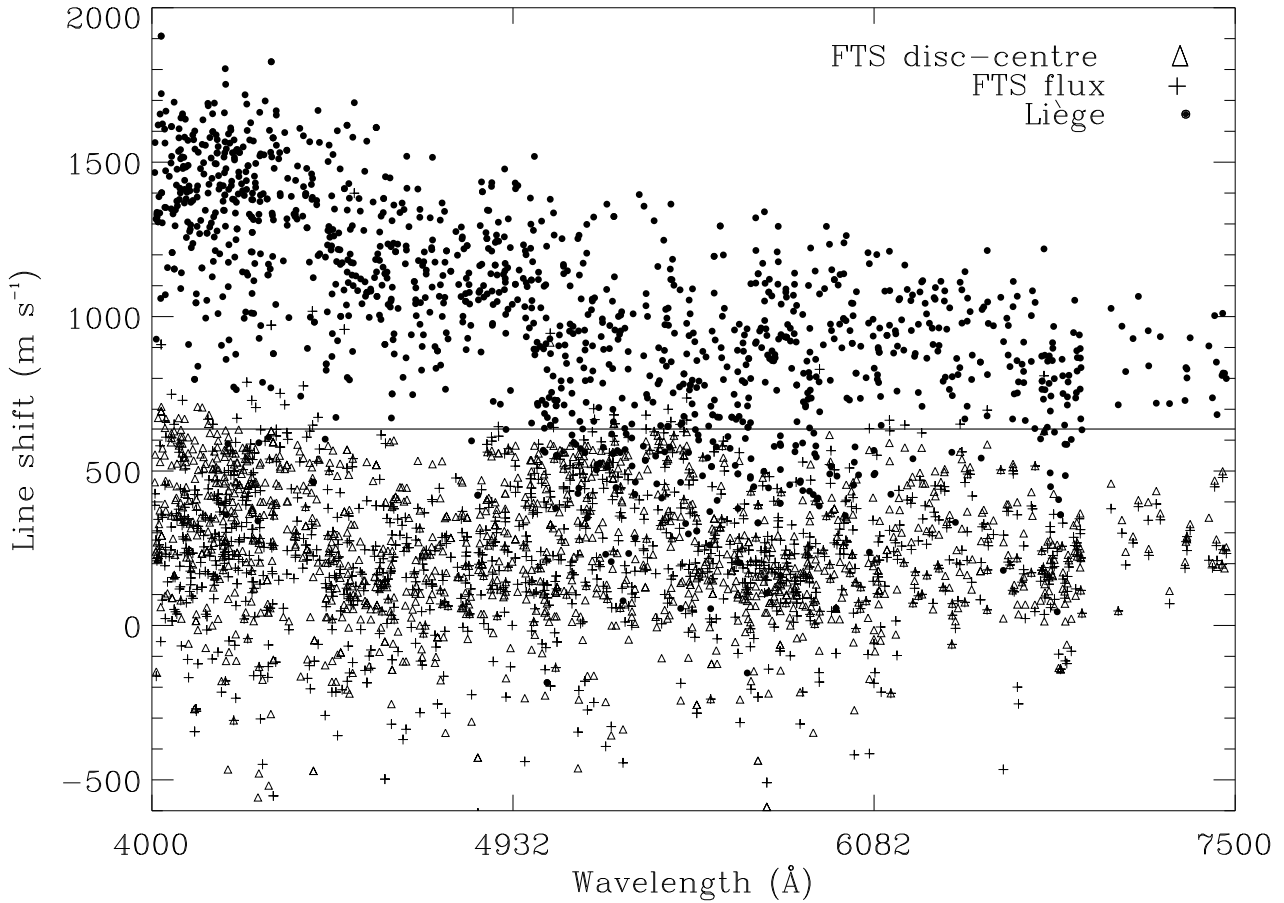,height=13.cm,angle=90}}
\caption{Line shifts measured in the FTS disc-centre spectrum (triangles)   and
in the FTS flux spectrum (plus signs) show no trend with wavelength.
The largest shifts to the red are consistent with the expected gravitational
shift of 636 m s$^{-1}$ (solid line). The shifts measured in the Li\`ege Atlas (dots) exhibit a wavelength dependence, and go further to the
red than the gravitational redshift, thereby revealing calibration errors.}
\end{center}
\end{figure*}

\section{Absolute wavelength calibration of the atlases}
\label{sec2}

As a first step, the wavelength calibrations of the two FTS atlases
mentioned above were tested. This was also done for the Li\`ege 
Atlas, which is composed of  disc-centre observations performed at
Jungfraujoch Observatory (Swiss Alps) using a double-pass grating
spectrometer, and which has been extensively used in many solar studies.
We have made use of the version available as part of the
KIS\footnote{Kiepenheuer-Institut f\"ur Sonnenphysik, Freiburg,
Germany.} IDL library. 

It has been shown (e.g. Balthasar 1984) that once
the Sun-Earth velocity shifts have been corrected for, the maximum
displacement to the red exhibited by solar spectral lines
corresponds to the gravitational redshift due to the difference in
gravity between the solar  and terrestrial surfaces. This is the expected case for the three atlases under consideration, which use wavelengths on standard air (dry, 15$^{\rm o}$C, 760 mmHg). Solar-Earth doppler shifts  were corrected in the FTS flux spectrum, and the same is true for the tables of solar wavelengths prepared by Pierce \& Breckenridge (1973) at Kitt Peak, which were employed to calibrate the wavelength scale of the FTS disc-centre spectrum. Both FTS spectra were compiled from eight (flux) or seven (disc-centre) carefully overlapped scans. The Li\`ege atlas is supposed to be calibrated following the same reference, but prior to publication.

Following the procedure described in the next section, shifts for
neutral iron lines were measured in the atlases. Figure 1 shows the
results as a function of wavelength. It appears that both the FTS disc-centre spectrum  and the FTS flux spectrum agree in the absolute scale within the errors  and the
expected differences between intensity and flux measurements.  They do
not show any stronger than expected  trend (Hamilton 1997) and are in agreement with the expected maximum
redshift of 636 m s$^{-1}$ (solid line), which corresponds to the
gravitational effect.  On the contrary, the wavelength calibration of
the Li\`ege Atlas differs clearly on the absolute scale, exhibiting
redshifts larger than 636 m s$^{-1}$ and showing a strong trend with
wavelength, which points towards errors in the spectral calibration
procedure. For this reason we have discarded the Li\`ege Atlas in this
study.

\section{Line shifts}
\label{sec3}

\subsection{Intensity}

From the most extensive  up-to-date measurements of solar wavelengths by
Pierce \& Breckenridge (1973),   DLN extracted a list of 311
unblended Fe\,{\sc i} lines. After subtracting the gravitational redshift,
they studied the trend of the shifts with the excitation potential of the
lines and placed the bisectors measured in the solar atlas of Li\`ege Atlas on an absolute scale. Their study demonstrated the power of line
asymmetries and shifts in providing a deeper understanding of solar convection.

A version of the FTS disc-centre spectrum, which has been
interpolated to a constant wavelength step, can be obtained as part of the KIS IDL library. We have measured on this atlas the  location of the minima
of 1446 Fe\,{\sc i} lines selected from the list of Th\`evenin (1989, 1990), thereby
extending the existing data provided by DLN and improving its accuracy significantly. Th\`evenin's list includes 2536 lines of neutral iron. Although
many of the lines are blended, only those showing clear evidence for the
blend to disturb the line centre were rejected. A fourth-order polynomial was
fitted to the 50 m\AA\ wavelength interval around the line bottom to find the line centre as precisely as
possible. Measured central wavelengths, together with their
corresponding air values at rest, excitation potentials, transition
probabilities and equivalent widths, are listed in  Table 1. Since the measurement process was automatic, Table 1 may contain certain errors for the observed wavelengths.

The centres of those lines formed higher in the photosphere
show smaller blueshifts. It can
be seen in Figure 2a that even those lines located in a ``plateau''
with equivalent widths ($W_{\lambda}$) larger than $\sim$ 100 m\AA\ seem to be somewhat
blueshifted, and there are no lines where the effect is negligible. The
mean\footnote{Only lines without an asterisk in the equivalent widths shown
in Table 1 were included.} value for the shift of these lines is 539 m
s$^{-1}$ (subtracting the gravitational redshift of 636 m s$^{-1}$, the
minimum convective blueshift will be 97 m s$^{-1}$) with a standard
deviation of 86 m s$^{-1}$. No other source in the  literature,
including spectra and wavelength measurements, is  accurate enough to
perform a more reliable comparison discarding minor systematic effects
of the wavelength calibration.

\subsection{Flux}

To the best of our knowledge, the only existing measurements of line shifts
from solar flux spectra are those of Burns, Meggers, and Kiess, published in 1929 (Burns 1929; Burns \& Kiess 1929; Burns \& Meggers 1929). Nonetheless, systematic analyses of shifts and asymmetries of spectral
lines in the Sun seen as a star are of special importance as a
reference standard for comparison with other stars.

The FTS flux spectrum is available from the
NOAO\footnote{National Optical Astronomical Observatories, USA.} ftp
site. We have measured the wavelengths of the minima of the same 1446
Fe\,{\sc i} lines considered in the FTS disc-centre spectrum. Similarly to
the intensity case, a fourth-order polynomial was fitted to the 55 m\AA\ wavelength interval around the line centre. The results are included in the second
column of Table 1. This list provides the largest available set of
accurate solar wavelengths useful as a reference for comparison with
other solar-type stars.

The smoothing of the convective blueshift towards the solar limb, the
so-called limb effect, results in a smoothing of the convective shifts
in the integrated sunlight compared with the intensity spectrum. In
this atlas, the ``plateau'' with  equivalent widths larger that $\sim$
200 m\AA\ is formed by lines  distributed around the  gravitational
redshift, as  can be seen in Figure 2b. It makes sense that the
smoothing of the blue-shifts at the disc centre results in a plateau
closer to  a null velocity shift. The scatter (standard deviation) in
these line shifts at the plateau is 58 m s$^{-1}$, around a mean value
of 612 m s$^{-1}$ (the subtraction of the gravitational redshift comes
with a mean of 24 m s$^{-1}$ for the smaller convective blueshifts).

Wallace, Huang \& Livingston (1988) studied the variability of the convective line shifts in the solar flux spectrum during the solar cycle measuring the relative shifts between a weak line, C \,{\sc i} $\lambda$5380.3 \AA\ ($W_{\lambda}$ = 26 m\AA\ ) and the stronger features Fe \,{\sc i} $\lambda$5379.6 \AA\ ($W_{\lambda}$ = 67 m\AA\ ) and Ti \,{\sc ii} $\lambda$5381.0 \AA\  ($W_{\lambda}$ = 70 m\AA\ ). They found an upper limit of 5 m s$^{-1}$ for the relative line shift and concluded with the possibility of detecting Jupiter, which would produce a 20 m s$^{-1}$ amplitude in the solar radial velocity,  from an extra-solar system observer. However, Figure 2b suggests that the $\sim$ 70 m\AA\ pair of lines they employed as reference wavelengths less affected by convection were not the ideal choice. Comparison between a much stronger feature and the C\,{\sc i} line will probably give a safer answer on this subject, where different measurement techniques observe (Deming \& Pymate 1994) and deny (McMillan et al. 1993) the variations of the convective shifts.

\begin{figure*}
\begin{center}
\mbox{\epsfig{file=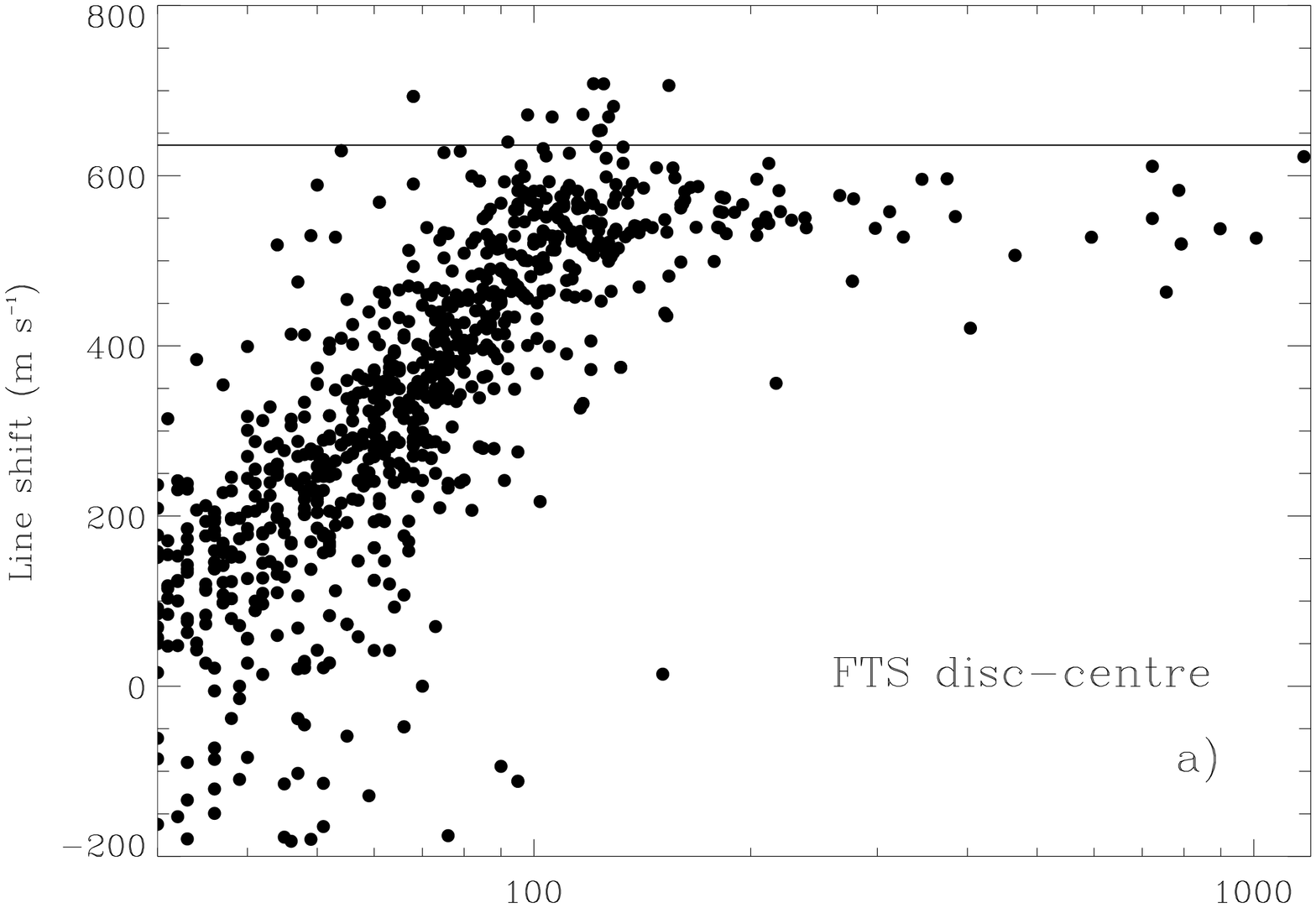,height=10.cm,angle=90}}
\mbox{\epsfig{file=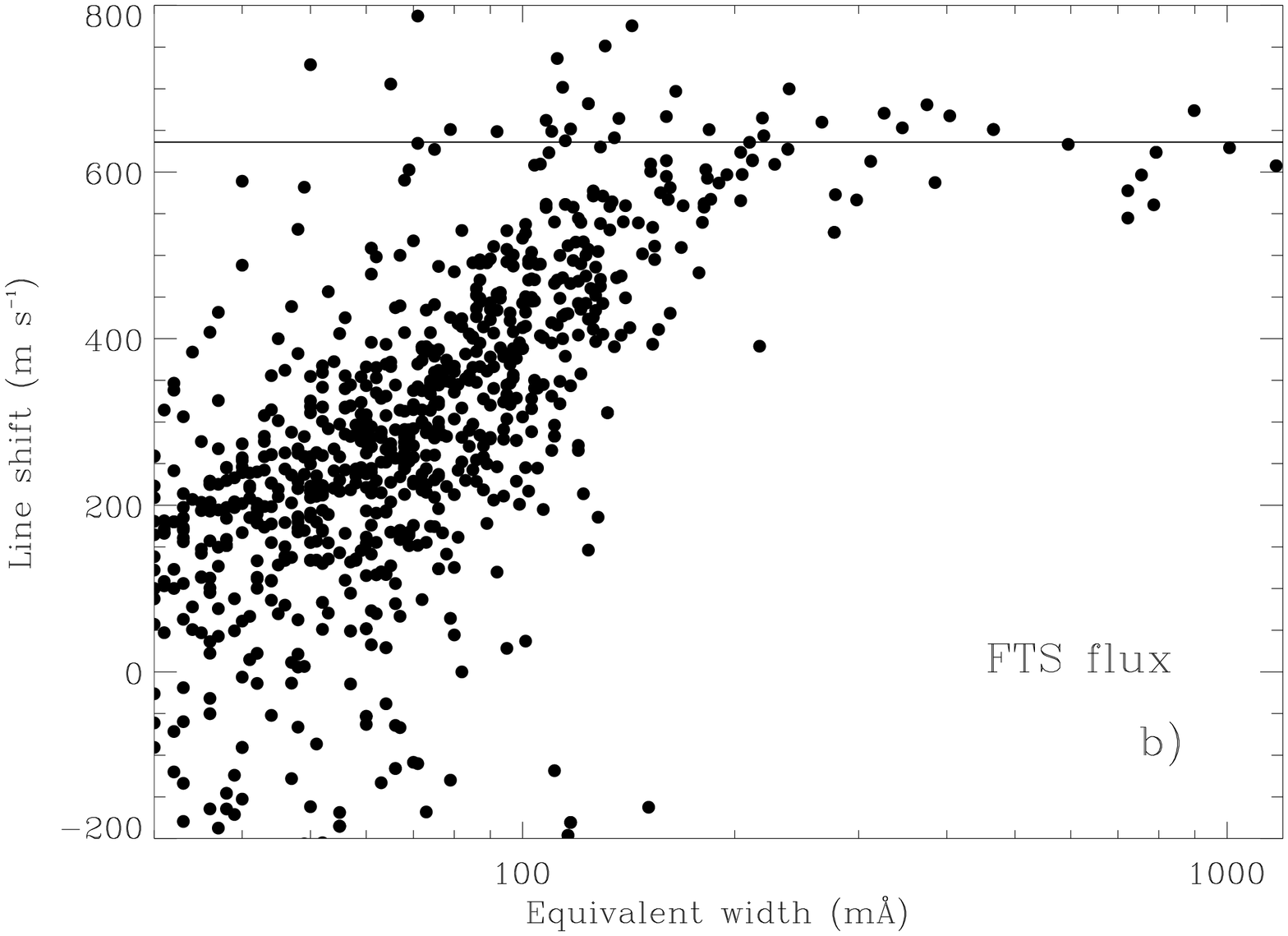,height=10.cm,angle=90}}
\end{center}
\caption{Line shifts as measured in the  FTS disc-centre spectrum (a) and the FTS flux spectrum (b) are plotted as a function of the  equivalent 
width at the centre of the disc (Moore et al. 1966). Known blends (marked with an asterisk in
Table 1) have been rejected.}
\end{figure*}

\newpage

\section{Conclusions}
\label{sec4}

We have verified that the line wavelength shifts  measured in the FTS
disc-centre spectrum and the FTS flux spectrum agree on an absolute
scale and do not show any strong trend with wavelength, whereas the
Li\`ege Atlas exhibits an anomalous wavelength dependence with line
shifts in excess of the gravitational redshift, thereby revealing
calibration errors.

Accurate wavelengths have been measured for 1446 Fe\,{\sc i} lines, both from
intensity and flux spectra, providing the largest database available for
comparison with other solar-type stars. This line list may be useful in
employing solar (daylight or lunar) spectra to perform wavelength calibrations
in high-resolution spectroscopy, or to assess the quality of calibrations based on spectral lamps, which normally illuminate the detector through a different optical path from that of the observational target.

Finally, it has been shown how the wavelength shifts of lines formed at
the top of the photosphere get close to the gravitational redshift.

\acknowledgements{We thank H. Neckel for the help  in dealing with his
solar atlas,  F. Th\`evenin and A. D. Wittmann for providing us with a
copy of their digital solar line lists, and H. H. R. Kroll for
installing and maintaining the KIS computer libraries at the IAC. We
are grateful to J. S\'anchez Almeida for interesting comments after 
careful reading of the manuscript.  Valuable bibliographic information was pointed out to us by the referee. NSO/Kitt Peak FTS data used here
were produced by NSF/NOAO.

This work was partially supported by the Spanish DGES under projects
PB92-0434-C02-01 and PB95-1132-C02-01.}

\end{document}